\documentclass[aps,prd,preprint,superscriptaddress]{revtex4}

\usepackage{graphicx}
\usepackage{amsmath} 
\usepackage{amssymb}
\usepackage{amsfonts} 
\usepackage{bm}
\usepackage[usenames]{color}

\def\ee{\end{eqnarray}}
\def\Lag{\mathcal{L}}
\def\p{\partial}

\def\=:{=\hspace{-.7em}\raisebox{1.1ex}{.}\hspace{.1em}\raisebox{-0.2ex}{.} }

\def\ee{\end{eqnarray}}
\def\Lag{\mathcal{L}}

\def\p{\partial}

\def\=:{=\hspace{-.7em}\raisebox{1.1ex}{.}\hspace{.1em}\raisebox{-0.2ex}{.} }

\newcommand {\beq}{\begin{eqnarray}}
\newcommand {\eeq}{\end{eqnarray}}
\newcommand {\non}{\nonumber\\}

\newcommand {\Sig}{\Sigma}

\newcommand{\SU}{SU}

\newcommand{\U}{U}
\def\Tr{{\rm Tr}}
\def\Lag{\mathcal{L}}

\newcommand\diag{\operatorname{diag}}
\def\Tr{\mathop{\mathrm{Tr}}}

\def\cut#1{\textcolor{Gray}{{ #1}}}

\begin{document}


\title{
Domain walls and vortices in chiral symmetry breaking
}


\author{Minoru Eto}
\affiliation{Department of Physics, Yamagata University, Yamagata
990-8560, Japan
}
\author{Yuji Hirono}
\affiliation{
Department of Physics, University of Tokyo, 
Hongo~7-3-1, Bunkyo-ku, Tokyo 113-0033, Japan
}
\affiliation{
Theoretical Research Division, Nishina Center, RIKEN, Wako 351-0198,
 Japan
}
\author{Muneto Nitta}
\affiliation{
Department of Physics, and Research and Education Center for Natural 
Sciences, Keio University, Hiyoshi 4-1-1, Yokohama, Kanagawa 223-8521,
Japan
}


\date{\today}
\begin{abstract}
We study domain walls and vortices  in
chiral symmetry breaking in a QCD-like theory with $N$ flavors in the chiral limit.
If the axial anomaly is absent, 
there exist stable 
Abelian axial vortices winding around the spontaneously broken 
$U(1)_{\rm A}$ symmetry
and non-Abelian axial vortices 
winding around both 
the $U(1)_{\rm A}$ and non-Abelian $SU(N)$ chiral symmetries. 
In the presence of the axial anomaly term, 
metastable domain walls are present  
and Abelian axial vortices must be attached by 
$N$ domain walls, forming domain wall junctions. 
We show that a domain wall junction decays 
into $N$ non-Abelian vortices attached by 
domain walls,  
implying its metastability. 
We also show that domain walls decay 
through the quantum tunneling 
by creating 
a hole bounded by a closed non-Abelian vortex.  
\end{abstract}
\pacs{}

\maketitle

\section{Introduction}

Domain walls produced at phase transitions are known to cause a conflict
with cosmology. 
When the Universe undergoes a phase transition, 
domain walls are formed 
if the order parameter space allows them. 
These domain walls dominate the energy density of the Universe, which
is not acceptable from a cosmological point of view (the domain wall problem).
Phase transitions at very high energies 
($\sim 10^{16}{\rm GeV}$) are
not harmful, since the energy density is diluted by the inflation.
However, phase transitions below that scale can be dangerous. 
It is well known that axion models, 
which are elegant extensions of the standard model for solving the
strong CP problem
\cite{Peccei:1977hh,Peccei:1977ur,Weinberg:1977ma,Wilczek:1977pj}, 
suffer from this problem if the number of flavors is larger than one 
\cite{Sikivie:1982qv, Peccei:1977hh,Peccei:1977ur,Dine:1981rt}.

The chiral phase transition in quantum chromodynamics(QCD) is apparently problematic 
as pointed out in Ref.~\cite{Forbes:2000et},  
since domain walls may be produced at the chiral symmetry breaking
\cite{Zhang:1997is,Balachandran:2001qn,Balachandran:2002je}. 
It is still unclear whether the domain walls actually form or not in the early Universe, because
the chiral symmetry breaking is a crossover  
as a function of temperature 
rather than a phase transition at zero baryon density. 
If the crossover is very dull, no production of the domain walls is possible.
In contrast, the domain walls are expected to form if the crossover is sharp enough and
the chiral condensation rapidly grows. In order to clarify this point, one must examine
the relaxation timescales involved, which are beyond the scope of this work.
In the latter case,
in particular, a junction of three domain walls 
glued by an Abelian axial vortex was found 
in Ref.~\cite{Balachandran:2001qn} so that 
domain wall network would be produced 
that would make domain walls long-lived.  
Also, a heavy-ion collider at GSI is designed to achieve 
a finite baryon density, 
which may turn the chiral symmetry breaking from a crossover to a sharp transition. 
In that case, the production of topological defects will be 
inevitable.

In this paper, 
we study the (in)stability of domain walls 
and vortices in chiral symmetry breaking 
in a QCD-like theory 
in which we take into account light scalar mesons 
while ignoring other heavy modes such as vector mesons. 
We find that domain wall junctions are metastable and 
decay into separate multiple domain walls 
edged by non-Abelian axial vortices \cite{Balachandran:2002je},  
which are the fundamental vortex solutions 
\cite{Nitta:2007dp,Nakano:2007dq,Eto:2009wu}.  
We show that the decays are possible from a topological point of view 
and perform numerical simulations of decaying junctions. 
We next show that domain walls themselves can decay by making use of
non-Abelian vortices. 
In a domain wall, holes whose boundaries are 
non-Abelian vortices can be 
excited quantum-mechanically or thermally. 
We make an estimate of the decay rate. 

The same types of topological defects also exist 
in QCD at high baryon density \cite{Eto:2013hoa}, 
where the color-flavor locked phase is realized 
and the chiral symmetry is spontaneously broken
\cite{Alford:1997zt,Alford:1998mk,Alford:2007xm}.
The chiral Lagrangian in this case was discussed in 
Ref.~\cite{Casalbuoni:1999wu}. 
Therefore, the same discussions in this paper hold 
also for high-density QCD.

This paper is organized as follows. 
In Sect.~\ref{sec:GL}, we present the Ginzburg-Landau 
effective theory for the chiral symmetry breaking. 
In Sect.~\ref{sec:wall-vortex}, we describe 
domain walls and vortices. 
In Sect.~\ref{sec:wall-vortex-composite}, 
we consider composite states of domain walls and vortices 
in the presence of the axial anomaly term. 
We numerically construct a three-domain wall junction 
for three flavor QCD. 
In Sect.~\ref{sec:instability}, we show the instability of 
the domain wall junction topologically and 
simulate such a decay numerically.  
Section \ref{sec:summary} is devoted to the summary 
and discussion. 

\section{Ginzburg-Landau effective Lagrangian}\label{sec:GL}

The chiral symmetry 
$SU(N)_{\rm L} \times SU(N)_{\rm R} \times U(1)_{\rm A}$ 
acts on $N$-flavor left- and right-handed massless chiral fermions 
$\psi_{{\rm L}i}$ and $\psi_{{\rm R}i}$ as
\beq
&& \psi_{{\rm L}i} \to e^{-i\theta_{\rm A}/2} g_{\rm L} \psi_{{\rm L}i}, 
\quad 
\psi_{{\rm R}i} \to e^{+i\theta_{\rm A}/2} g_{\rm R} \psi_{{\rm R}i}, \non
&& \left(e^{i\theta_{\rm A}},g_{\rm L},g_{\rm R}\right) \in U(1)_{\rm A} \times SU(N)_{\rm L} \times SU(N)_{\rm R}, 
\eeq 
where $U(1)_{\rm A}$ is explicitly broken by 
the axial anomaly.
When chiral condensation occurs, 
\beq
 \Sigma_{ij} \sim \left< \bar \psi_{{\rm L}i} \psi_{{\rm R}j}\right> 
\neq 0,
\eeq
the chiral symmetry is spontaneously broken. 
Here $\Sigma$ is an $N\times N$ complex matrix scalar field, transforming under the chiral symmetry as 
\beq
\Sigma \to e^{i\theta_{\rm A}} g_{\rm L}^\dagger \Sigma g_{\rm R},\qquad
\left(e^{i\theta_{\rm A}},g_{\rm L},g_{\rm R}\right) \in U(1)_{\rm A} \times SU(N)_{\rm L} \times SU(N)_{\rm R}.
\eeq 
There is a redundancy in the chiral symmetry acting on the scalar field $\Sigma$. 
The true symmetry group is written as 
\beq
G= \frac{SU(N)_{\rm L} \times SU(N)_{\rm R} \times U(1)_{\rm A}}{({\mathbb Z}_N)_{\rm L+A} \times ({\mathbb Z}_N)_{\rm R+A}} 
\simeq {SU(N)_{\rm L} \times SU(N)_{\rm R} \times U(1)_{\rm A} \over
  ({\mathbb Z}_N)_{\rm L+R} \times  ({\mathbb Z}_N)_{\rm L-R+A}},
\eeq 
where the redundant discrete groups are 
\beq
(\mathbb{Z}_N)_{\rm L+A} &:& 
\left(\omega_N^k{\bf 1}_N,{\bf 1}_N,\omega_N^{-k}\right)
 \in \SU(N)_{\rm L} \times \SU(N)_{\rm R} \times \U(1)_{\rm A},
\label{eq:ZN_L+A}\\
(\mathbb{Z}_N)_{\rm R+A} &:& 
\left({\bf 1}_N,\omega_N^{k}{\bf 1}_N,\omega_N^{-k}\right)
 \in \SU(N)_{\rm L} \times \SU(N)_{\rm R} \times \U(1)_{\rm A},
\label{eq:ZN_R+A}\\
(\mathbb{Z}_N)_{\rm L+R} &:& 
\left(\omega_N^k{\bf 1}_N,\omega_N^{-k}{\bf 1}_N,1\right)
 \in \SU(N)_{\rm L} \times \SU(N)_{\rm R} \times \U(1)_{\rm A},
\label{eq:ZN_L+R}\\
(\mathbb{Z}_N)_{\rm L-R+A} &:& 
\left({\omega_N^k\bf 1}_N,\omega_N^{k}{\bf 1}_N,\omega_N^{-2k}\right)
 \in \SU(N)_{\rm L} \times \SU(N)_{\rm R} \times \U(1)_{\rm A},
\label{eq:ZN_L-R+A}  
\eeq
with $\omega_N \equiv e^{i\frac{2\pi}{N}}$ 
and $k=0,1,\cdots,N-1$.

The generic Ginzburg-Landau effective Lagrangian $\Sigma$ in the chiral
limit 
can be written as \cite{Pisarski:1983ms}
\begin{eqnarray}
 \Lag &=& \Tr\left[\p_\mu \Sigma^\dagger \p^\mu \Sigma 
- \lambda_2(\Sigma^\dagger \Sigma)^2 +
\mu^2 \Sigma^\dagger \Sigma\right] 
- \lambda_1 \left(\Tr\left[\Sigma^\dagger\Sigma\right]\right)^2 
+ C (\det \Sigma + {\rm c.c.}),
\label{eq:lag_lsm} 
\end{eqnarray}
where $\lambda_1$, $\lambda_2$, $\mu$, and $C$ are real parameters.
The last term in the Lagrangian (\ref{eq:lag_lsm}) is 
the axial anomaly term \cite{Pisarski:1983ms}, 
which breaks the $U(1)_{\rm A}$ symmetry explicitly. 
In this paper, we consider the chiral limit in which 
all the quarks are massless.

Note that we do not 
take into account other massive fields,  
which are possibly light at high temperature or high baryon density,
such as vector mesons and baryons. 
Since we are interested in topological defects 
at the chiral symmetry breaking, 
all the essential points can be extracted from the Ginzburg-Landau theory in Eq.~(\ref{eq:lag_lsm}).
One can refine the analysis in this paper 
by taken into account all the fields,
although the results would not be unchanged
qualitatively.

We consider the phase in which the chiral symmetry is spontaneously broken, so we assume that the constants in Eq.~(\ref{eq:lag_lsm}) satisfy the relations $\mu^2 > 0$ and $N\lambda_1 + \lambda_2 > 0$ for the  vacuum stability. 
One can choose the ground state value as
\beq
\Sigma = v {\bf 1}_N,\qquad
v \equiv \sqrt{\frac{\mu^2}{2(N\lambda_1 + \lambda_2)}} ,
\eeq
for $C=0$ without loss of generality. 
In the ground state, the chiral symmetry $G$ is spontaneously broken down to its diagonal subgroup
\beq
 H
= 
 {SU(N)_{\rm L+R} \times ({\mathbb Z}_N)_{\rm L-R+A}\over 
({\mathbb Z}_N)_{\rm L+A} \times  ({\mathbb Z}_N)_{\rm R+A}}
\simeq {SU(N)_{\rm L+R} \over ({\mathbb Z}_N)_{\rm L+R}}.
  \label{eq:HF}
\eeq
This spontaneous symmetry breaking results in  
$N^2-1$ $SU(N)$ Nambu-Goldstone (NG) particles in addition 
to a $U(1)_{\rm A}$ NG particle.  
The $U(1)_{\rm A}$ symmetry is explictly broken by the axial anomaly 
so that the corresponding particle is a pseudo-NG boson, 
which we shall call the $\eta'$ meson.
The mass spectra are as follows: there are $N^2$ massive bosons
whose masses are
\beq
m_1^2 = 2\mu^2,\quad
m_{\rm adj}^2 = 4\lambda_2v^2,
\eeq
for the components in singlet and adjoint representations of $SU(N)_{\rm L+R}$, respectively. 

When $C > 0$,  $\eta'$ gets a finite mass, 
\beq
 m_{\eta'}^2 = CNv^{N-2},
\label{eq:eta-mass}
\eeq
and the order parameter space reduces as
\beq
{G \over H} 
\simeq  
U(N)_{\rm L-R+A}
\quad \xrightarrow{\ \ C\neq 0\ \ } \quad
SU(N)_{\rm L-R}.
\label{eq:GFHF}
\eeq

Let us assume that $m_{\eta'}$ is much smaller than $m_1$ and $m_{\rm adj}$ which is likely to occur
at high temperature or high baryon density, 
where instanton effects are suppressed; namely, we assume that $C$ is sufficiently small.
Then, we can integrate out the heavier
fields with the masses $m_1$ and $m_{\rm adj}$, so that
the Lagrangian (\ref{eq:lag_lsm}) reduces to a nonlinear sigma model 
(the chiral Lagrangian). This can be easily verified as follows.
Since the coupling constant $C$ in the effective Lagrangian (\ref{eq:lag_lsm})
is much smaller than the others, we can fix the amplitude of $\Sigma$ as
\beq
\Sigma =  v e^{i\varphi_{\rm A}} U,\quad UU^\dagger = {\bf 1}_N,
\label{eq:restrict}
\eeq
where the $U(1)_{\rm A}$ Nambu-Goldstone mode $\eta'$ takes a value in $\varphi_{\rm A} \in [0,2\pi)$.
Plugging this into Eq.~(\ref{eq:lag_lsm}), one gets an
effective Lagrangian for the mesons: 
\begin{equation}
\Lag_{\rm eff} 
= 
v^2 \Tr\left[ \p_\mu U \p^\mu U^\dagger\right] + v^2 N  \p_\mu\varphi_{\rm A}\p^\mu\varphi_{\rm A}
+ 2 v^NC \cos N\varphi_{\rm A}.
\label{eq:axion_3}
\end{equation}
It is straightforward to read the $\eta'$ mass in Eq.~(\ref{eq:eta-mass}) from this.
The Lagrangian (\ref{eq:axion_3}) is nothing but the sine-Gordon model
with a period
$\varphi_{\rm A} \sim \varphi_{\rm A} + 2\pi/N$. 
There exist $N$ discrete vacua in the $U(1)_{\rm A}$ space:
\beq
\varphi_A = (\omega_N)^a,\quad 
\qquad(a=0,1,2,\cdots,N-1).
\label{eq:N_vac}
\eeq

\section{Domain walls and vortices}\label{sec:wall-vortex}

The phase with the broken chiral symmetry 
accommodates metastable
domain walls and vortices. 
We discuss domain walls and vortices
in the first and second subsections, respectively. 

\subsection{Domain walls}
\label{sec:dw}

The Lagrangian (\ref{eq:axion_3}) allows domain wall solutions 
\cite{Skyrme:1961vr}, which interpolate two adjacent vacua among the $N$ vacua.
One minimal-energy configuration is a domain wall 
that interpolates between $\varphi_{\rm A} = 0$ at $x = -\infty$ 
and $\varphi = 2\pi/N$ at $x=\infty$. 
Assuming that the field depends only on 
one space direction, say $x$, 
an exact solution of a single static domain wall 
can be obtained as 
\beq
\varphi_{\rm A} (x) = \frac{4}{N} \arctan e^{m_{\eta'}
 (x-x_0)}, \label{eq:fractional-SG}
\eeq
where  $x_0$ denotes the position 
of the domain wall. 
The tension of the domain wall is given by
\beq
T_{\rm w} = \frac{16}{N}v^2m_{\eta'}.\label{eq:tension-wall}
\eeq
A typical scale of the domain wall is
\beq
\ell_{\rm dw} = m_{\eta'}^{-1}.
\eeq

The other $N-1$ minimal domain walls  
are simply obtained by shifting the phase as $\varphi_A \to \varphi_A + 2\pi a/N$ ($a=1,2,\cdots,N-1$).
The anti-domain walls are also easily obtained just by reflection $x \to -x$.
All of the domain walls wind 
the $U(1)_{\rm A}$ phase $1/N$ times,   
unlike the unit winding for the usual sine-Gordon domain walls. 
Therefore, we call these domain walls as 
{\it fractional} axial (sine-Gordon) domain walls.
Two fractional sine-Gordon domain walls 
repel each other 
(the repulsion $\sim e^{-2R}$ with distance $2R$) 
\cite{Perring:1962vs}.

The existence of the domain walls is obvious from the above discussion. 
However, note that the $N$ vacua given in Eq.~(\ref{eq:N_vac}) are not discrete but 
are all continuously connected via the $SU(N)_{\rm L-R}$ space.
To see this, let us consider the $N=3$ case as a simple example. Let us introduce two paths inside $SU(3)_{\rm L-R}$ as
\beq
P_1(\alpha) = \left(
\begin{array}{ccc}
1 &&\\
& e^{i\alpha} &\\
& & e^{-i\alpha}
\end{array}
\right),\quad
P_2(\alpha) = \left(
\begin{array}{ccc}
e^{-i\alpha} &&\\
& 1 &\\
& & e^{i\alpha}
\end{array}
\right),
\eeq
with $\alpha \in [0,2\pi/3]$. The two vacua $\left<\Sigma\right>_1 = v {\bf 1}_3$ and 
$\left<\Sigma\right>_2 = v \omega_3{\bf 1}_3$
are transformed as
\beq
\left<\Sigma\right>_1 &\to& P_1(\alpha) \left<\Sigma\right>_1 P_1(\alpha) = 
{\rm diag}\left(1, e^{2i\alpha},e^{-2i\alpha}\right),\\
\left<\Sigma\right>_2 &\to& P_2(\alpha) \left<\Sigma\right>_2 P_2(\alpha) = 
\omega_3 {\rm diag}\left(e^{-2i\alpha}, 1,e^{2i\alpha}\right).
\eeq
When $\alpha = 2\pi/3$, both $\left<\Sigma\right>_1$ and $\left<\Sigma\right>_2$ 
become $\left(1,\omega_3,\omega_3^2\right)$.
From this concrete example, it is obvious that there exist continuous
paths inside $SU(N)_{\rm L-R}$ that 
connect any two of the $N$ vacua given in Eq.~(\ref{eq:N_vac}). Since there are no potential barriers 
along the $SU(N)$ paths, it is possible to connect two vacua, say $\varphi_{\rm A} = 0$ and $\varphi_{\rm A} =\omega_N$
without any domain walls. Such a configuration costs only kinetic energy whose density is roughly $\sim v^2/L^2 \to 0$ as
$L \to \infty$ ($L$ is the size of the system).

Whether a domain wall is produced or not depends on distribution of the vacua at the chiral phase transition.
If a path connecting two vacua goes inside the $U(1)_{\rm A}$ space, a domain wall is produced. But
if a path goes inside the $SU(N)_{\rm L-R}$ space, no domain walls are created.
One might suspect that probability of creating such a domain wall is zero since the number of paths going 
inside $SU(N)_{\rm L-R}$ is infinite while one going through $U(1)_{\rm A}$ is finite.  However,
as we will see below, appearance of domain walls is not rare, but they necessarily appear 
when vortices are created.
One might also suspect that the domain walls are unstable even locally. 
This is not the case: one can easily see that 
the domain walls are at least locally stable by examining small fluctuations around the domain wall background.
Since the $SU(N)_{\rm L-R}$ part and $U(1)_{\rm A}$ part are decoupled in Eq.~(\ref{eq:axion_3}),  
no tachyonic
instability can arise from the degrees of freedom of $SU(N)_{\rm L-R}$. Additionally, the degree of freedom $\varphi_A$
obeys the sine-Gordon Lagrangian which, as is well known, has no instability.
This is a sharp contrast to the pionic domain walls living inside $SU(N)$,
which are known to be locally unstable, see {\rm e.g.} Ref.~\cite{Son:2007ny}.

\subsection{Vortices in the absence of the axial anomaly}
\label{eq:axial_vortex}

Let us consider the case with $C=0$ throughout this subsection.
Note that the axial anomaly is always present in QCD independent of temperature,
so this subsection is provided for a pedagogical exercise.
The anomaly term $C$ will be taken into account in Sect.~\ref{sec:wall-vortex-composite}.
Stable topological vortices appear in this case
since the order parameter manifold 
$G_{\rm F}/H_{\rm F} \simeq \U(N)_{\rm L-R+A}$ is not simply connected,
{\it i.e.,} the first homotopy group is non-trivial, 
\beq
\pi_1 [\U(N)_{\rm L-R+A}] \simeq \mathbb{Z}.
\eeq 
In order to generate a non-trivial loop in 
the order parameter manifold, 
one may simply use $T_0 \sim {\bf 1}_N$ generator of $U(1)_{\rm A}$.
Such a loop corresponds to the $\eta'$ string \cite{Zhang:1997is,Balachandran:2001qn}
for which the order parameter  behaves as
\begin{equation}
\Sigma(r,\theta) \xrightarrow[]{r\to\infty} v \, 
e^{i\theta}\,{\bf 1}_N. 
\end{equation}
The $\eta'$ string is a kind of the global string and its tension is given by \cite{Nitta:2007dp}
\begin{equation}
T_{\U(1)_{\rm A}} = N \times 2\pi v^2 \log\frac{L}{\xi_{\rm a}} +
 \text{const.}\, , 
\end{equation}
with the size of the system $L$ and the size of the axial vortex 
$\xi_{\rm a} \sim m_1^{-1}$.
A typical scale of the axial $U(1)_{\rm A}$ vortex is
$\xi_{\rm a} \sim m_1^{-1}$.

However, the solution above is not a vortex with minimal energy.
One can construct a smaller loop inside the order parameter manifold by
combining the $U(1)_{\rm A}$ generator 
$T_0 \sim {\bf 1}_N$ and non-Abelian generators 
$T_a$ ($a=1,2,\cdots,N^2-1$) of $SU(N)_{\rm L-R}$ \cite{Balachandran:2002je}. 
This configuration is called the M$_1$ vortex \cite{Balachandran:2005ev,Eto:2013hoa}.
The typical configuration takes the form
\beq
\Sigma 
\xrightarrow[]{r\to\infty} 
v\,{\rm diag}\left(e^{i\theta},1,\cdots,1\right)
=v\,e^{i\frac{\theta}{N}} {\rm diag}\left(e^{i\frac{(N-1)\theta}{N}},e^{-i\frac{\theta}{N}},\cdots, e^{-i\frac{\theta}{N}}\right),
\label{eq:ansz_ngv}
\eeq
at far distances from the vortex core. 
From the right-hand side of Eq.~(\ref{eq:ansz_ngv}), 
one can infer that the corresponding loops 
wind $1/N$ of the $\U(1)_{\rm A}$ phase, 
and are generated by 
non-Abelian generators of $SU(N)_{\rm L-R}$ 
at the same time. 
They are called fractional vortices 
because of the fractional winding of the $\U(1)_{\rm A}$ phase, 
or non-Abelian vortices  
because of the contribution of the non-Abelian generators. 
The tension of a single non-Abelian axial vortex, which is proportional
to the winding number with respect to $U(1)_{\rm A}$ symmetry,
 is $1/N$ of that of an Abelian axial vortex \cite{Nitta:2007dp}:.
\begin{equation}
T_{U(N)_{\rm L-R+A}} = 2\pi v^2 \log\frac{L}{\xi_{\rm na}} + \text{const.}, \label{eq:tension-NAvor}
\end{equation}
with $\xi_{\rm na} \sim \min\{m_1^{-1},m_{{\rm adj}}^{-1}\}$.
A typical scale of the axial $U(1)_{\rm A}$ vortex is
\beq
\ell_{\rm U(N)_{\rm L-R+A}} \sim \min\{m_1^{-1},m_{{\rm adj}}^{-1}\}.
\eeq

The inter-vortex force at the leading order vanishes 
among vortices in different components \cite{Nakano:2007dq}.
A $U(1)_{\rm A}$ vortex can be marginally separated to
$N$ non-Abelian axial vortices as
\beq 
&& {\rm diag}\,(e^{i\theta},e^{i\theta},\cdots,e^{i\theta}) \non
&&\to {\rm diag}\,(e^{i\theta_1},1,1,\cdots) \times 
 {\rm diag}\,(1,e^{i\theta_2},1,\cdots) \times
\cdots
 {\rm diag}\,(1,\cdots,1,e^{i\theta_N}), \label{eq:U(1)A-decay}
\eeq  
at this order, 
where $\theta_{1,2,\cdots,N}$ denotes an angle coordinate  
at each vortex center.

\section{
Vortex-domain wall complex in the presence of the axial anomaly}
\label{sec:wall-vortex-composite}

Let us see how the instanton-induced potential, the last term in Eq.~(\ref{eq:lag_lsm}),
affects the vortices. So let us set $C > 0$ throughout this section.
As we have seen in Sec.~\ref{sec:dw}, the instanton-induced potential yields domain walls.
Vortices can also be produced when the approximate $U(1)_{\rm A}$ symmetry is spontaneously
broken at the chiral phase transition. 
Since the order parameter space is not $U(N)_{L-R+A}$ but $SU(N)_{\rm L-R}$ with a trivial first homotopy group
$\pi_1[SU(N)_{\rm L-R}]=0$, isolated vortices cannot exist 
but are accompanied by domain walls. 
Vortices are always attached by domain walls 
due to the instanton-induced potential, 
just as in the case of axion strings.

In the case of Abelian axial vortices, 
the phase changes from $\varphi_{\rm A}=0$ to
$\varphi_{\rm A} = 2\pi$ around a vortex.
Consequently,  
$N$ different domain walls forming an $N$-pronged junction attach to it.
The domain walls repel each other, 
so the configuration becomes a ${\mathbb Z}_N$-symmetric domain wall junction with
an Abelian vortex at the junction point.
A numerical solution for this configuration for $N=3$
was first obtained in Ref.~\cite{Balachandran:2001qn}. 
Here, we numerically reexamine 
the domain wall junctions.
As done in Ref.~\cite{Balachandran:2001qn}, 
we truncate the field as
\beq
\Sigma = \phi(x,y,t) {\bf 1}_N.
\eeq
We then obtain the reduced Lagrangian 
\beq
\Lag_{\rm red} = N |\p_\mu\phi|^2 - \frac{N m_1^2}{4v^2}\left(|\phi|^2 - v^2\right)^2 + C(\phi^N+\phi^{*N}).
\eeq
This can be rewritten in terms of the following dimensionless variables,
\beq
\phi \to v \phi,\quad x_\mu \to m_1 x_\mu,
\eeq
as
\beq
\Lag_{\rm red} = Nv^2m_1^2\left(|\p_\mu\phi|^2 - \frac{1}{4}\left(|\phi|^2 - 1\right)^2 + \frac{\tau}{N^2}(\phi^N+\phi^{*N})\right),
\quad
\tau\equiv \frac{m_{\eta'}^2}{m_1^2}.
\eeq
It is the dimensionless parameter $\tau$ that determines the properties of
the domain wall junctions.

\begin{figure}[ht]
\begin{center}
\includegraphics[width=16cm]{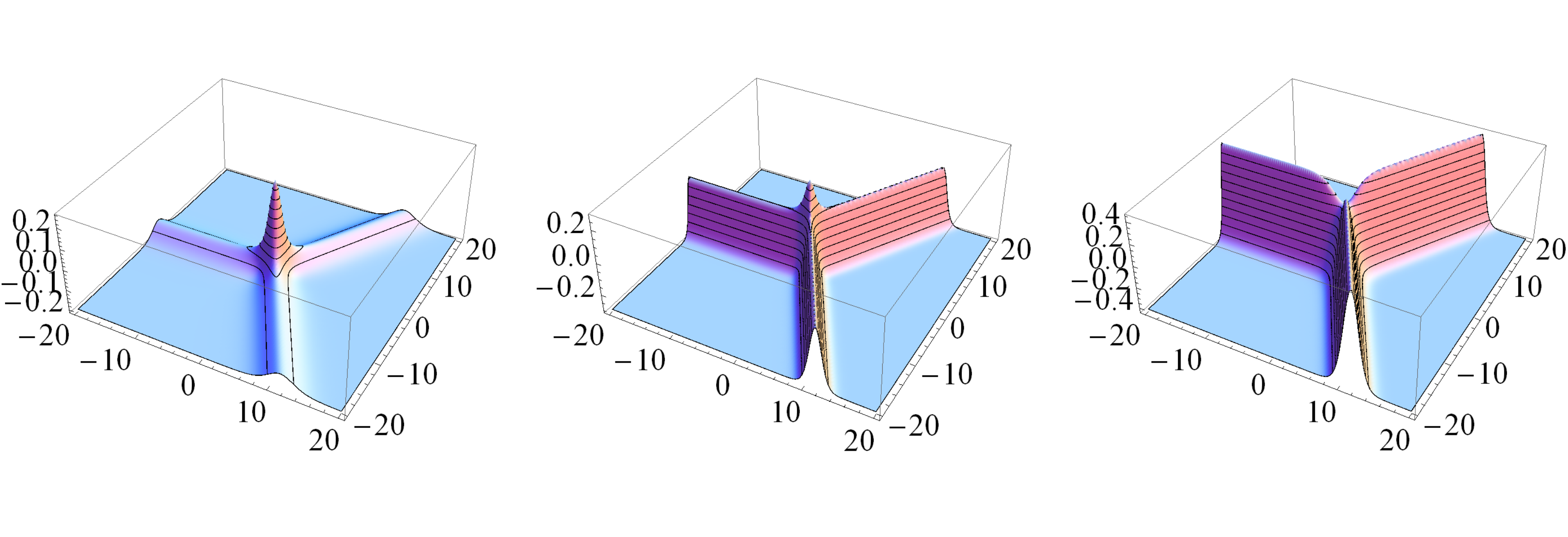}
\put(-450,120){(1)} 
\put(-295,120){(2)} 
\put(-140,120){(3)} 
\vspace*{-1cm}
\caption{Domain wall junctions in the $N=3$ case. 
The energy densities are plotted for different choices of the parameters:
(1) 
$\tau \simeq 0.1$ ($\lambda_1=\lambda_2=1$ and $C/\mu = 1/5$), 
(2) 
$\tau \simeq 0.5$ ($\lambda_1=\lambda_2=1$ and $C/\mu = 1$), 
(3) 
$\tau \simeq 12$ ($\lambda_1=\lambda_2=1/5$ and $C/\mu = 10$). 
The spatial axes are in units of  $m_1^{-1}$, and
the vertical axis is in units of $Nv^2m_1^2$.
}
\label{fig:junction}
\end{center}
\end{figure}

We make use of the so-called relaxation method to find static solutions;
namely we introduce an additional dissipative term in the equations of motion.
The scalar field $\phi$ obeys the following reduced equation of motion:
\beq
\ddot\phi + \gamma \dot \phi - \nabla^2 \phi = - \frac{\p V}{\p \phi^*} 
\label{eq:BB}
\eeq
where the dots denote differentiations with respect to time and 
the second term on the left-hand side 
is the dissipative term that we have introduced for the relaxation. 
In order to get an approximate numerical solution, we first solve the first-order
equation, which is obtained by discarding the second-order time derivative 
from Eq.~(\ref{eq:BB}).
The dissipative term 
deforms appropriate initial configurations 
and the configuration is converged
to the desired solutions, namely the domain wall junctions. 
Furthermore, in order to verify if the obtained solutions indeed satisfy the genuine field equation,
after the relaxation is done for sufficiently long period, one switches off the
dissipative term. Then, if the configurations do not evolve with the real time, it implies that
they are static and thus approximate solutions \cite{Balachandran:2001qn}.
We reproduced static solutions of the domain wall junctions, as shown in Fig.~\ref{fig:junction}.
Here, we show three examples with different relative tensions of the domain wall
and the Abelian axial vortex by changing the value of $\tau$. We find that the domain wall tension tends to be bigger (smaller)
than one of the vortices for bigger (smaller) $\tau$. 

Note that we should anticipate that the Abelian axial vortex might be broken up into three
non-Abelian axial vortices. However, the domain wall junctions cannot be broken up as long as
we work in the reduced model given in Eq.~(\ref{eq:BB}) since no non-Abelian vortices can be described
by the reduced equation of motion (\ref{eq:BB}).
In order to see if static domain wall junctions exist or not, 
we should leave more degrees of freedom 
\beq
\Sigma = {\rm diag}\left(\phi_1(x,y,t),\cdots,\phi_N(x,y,t)\right),
\label{eq:less_reduced}
\eeq
where the $N$ complex scalar fields are dealt with as independent fields.
In the case where no domain walls exist for $C=0$, 
well separated non-Abelian axial vortices experience no force at leading order \cite{Nakano:2007dq} 
and a repulsive force at the next leading order 
\cite{Eto:2011wp,Eto:2013hoa}, 
so that the Abelian axial vortex is not likely to be stable 
as in Eq.~(\ref{eq:U(1)A-decay}). 
Therefore, one would naively expect that there are no static domain wall junctions because
the Abelian axial vortex will be easily torn off into $N$ non-Abelian axial vortices
since the non-Abelian vortices are pulled by the domain walls toward different directions.  
Nevertheless, we found static domain wall junctions in the less-reduced models
with multiple complex scalar fields in Eq.~(\ref{eq:less_reduced}).
Several numerical solutions of static domain wall junctions are shown in Fig.~\ref{fig:junction}
for $N=3$ case.

Although we have found static solutions numerically, this does not
immediately imply their stability. 
In our case, they might be just stationary points of the action. Indeed, in the following sections, we will study disintegration of  Abelian axial vortices into  non-Abelian axial vortices.

\begin{figure}[ht]
\begin{center}
\includegraphics[height=4cm]{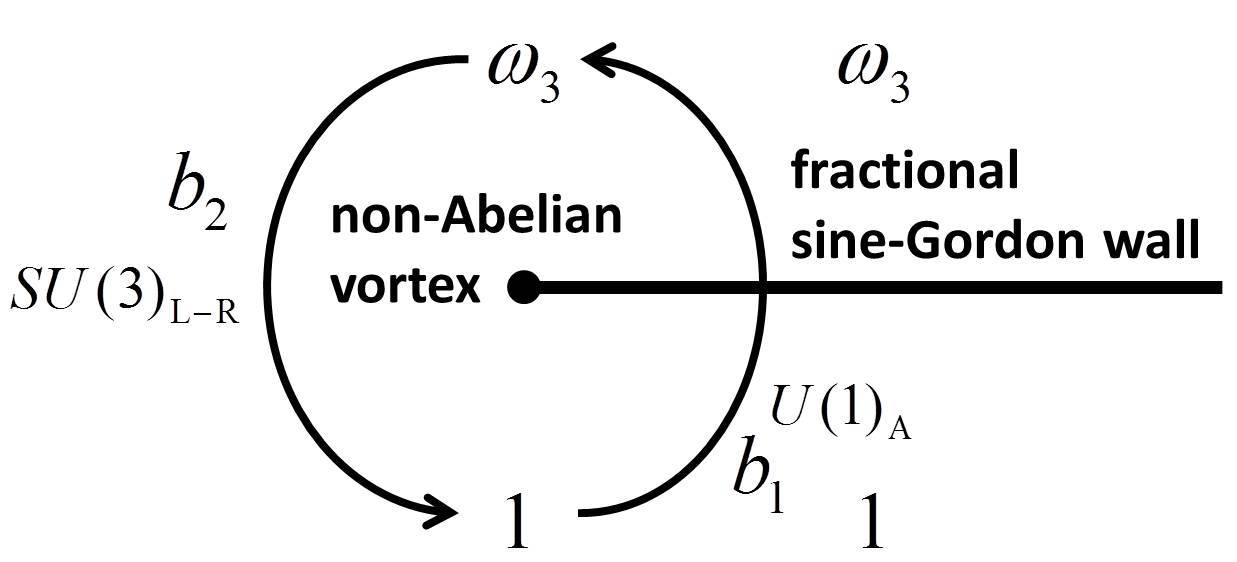}\\
(a)\\
\includegraphics[width=0.7\linewidth]{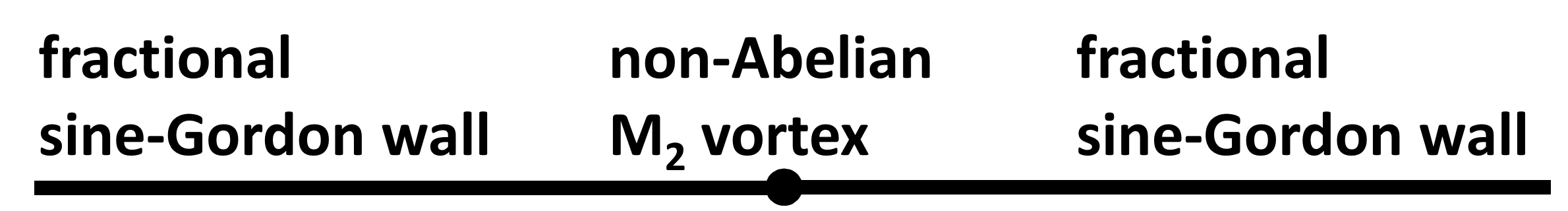}\\
(b)
\caption{
(a)
Non-Abelian axial vortex attached by a fractional axial 
domain wall. 
Along the path $b_1$, only the $U(1)_{\rm A}$ phase is rotated 
by $2\pi/3$. Then, the $SU(3)_{\rm L-R}$ transformation 
$\exp [( i /3) (\theta-\pi/2) \diag (2,-1,-1)]$ is performed 
along the path $b_2$, where $\theta$ ($\pi/2 \leq \theta \leq 3\pi/2$) is the angle of the polar coordinates  
at the black point.
(b) 
An M$_2$ non-Abelian axial vortex attached by two fractional axial domain walls. 
\label{fig:NAvortex-wall}
}
\end{center}
\end{figure}

\begin{figure}[ht]
\begin{center}
\includegraphics[width=17cm]{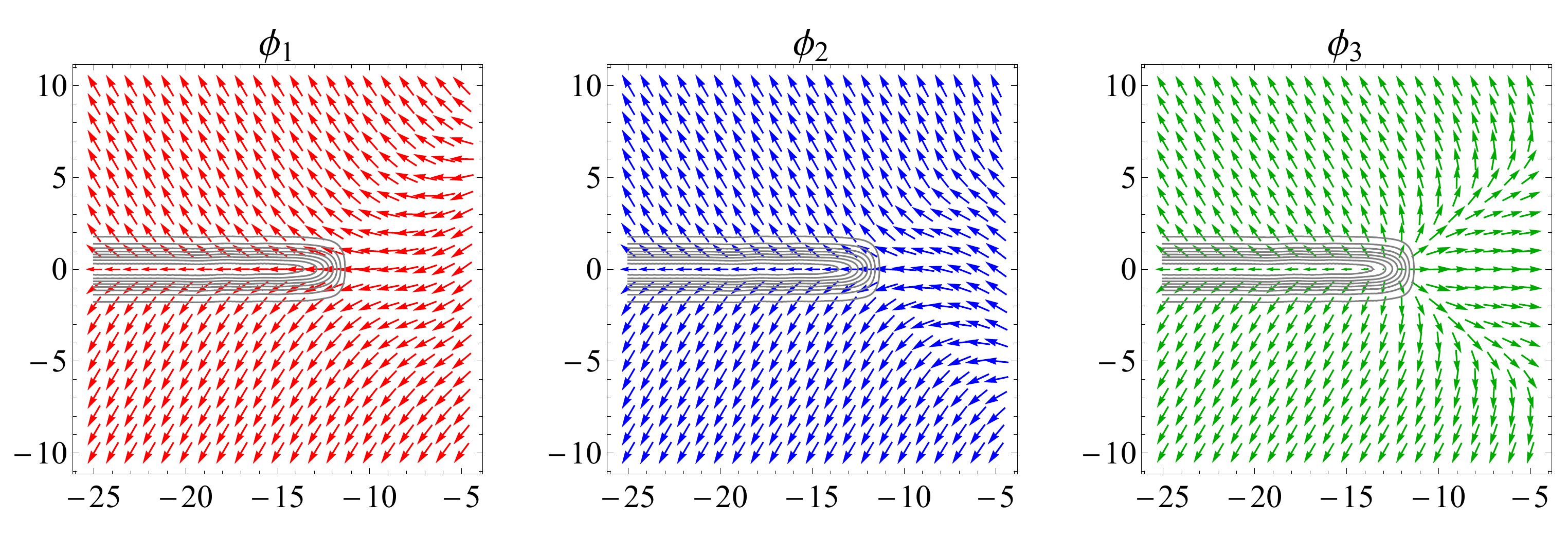}
\caption{The non-Abelian axial vortex attached by a fractional domain wall for $N=3$. 
The parameter is $\tau \simeq 0.5$ ($\lambda_1=\lambda_2=1$ and $C/\mu = 1$).
The non-Abelian axial vortex $(\phi_3)$  is located at $(x,y)\sim(-13,0)$ and
the domain wall extends toward the $-x$ direction from the vortex. 
The directions and the magnitudes  
of the arrows denote the phases of the amplitudes of 
$\phi_i$.
The spatial axes are in the unit of $m_1^{-1}$.
}
\label{fig:na}
\end{center}
\end{figure}

Let us next consider non-Abelian axial vortices. 
Since the $\U(1)_{\rm A}$ phase changes by $2\pi/N$ 
around a vortex, 
one fractional axial wall attaches to one non-Abelian axial vortex as illustrated in Fig.~\ref{fig:NAvortex-wall}(a).
Let us examine the structure in more detail, focusing on 
the configuration of the type $\diag (e^{i \theta},1,\cdots,1)$. 
In the vicinity of the vortex, 
let us divide a closed loop encircling the vortex 
into paths $b_1$ and $b_2$ as in 
Fig.~\ref{fig:NAvortex-wall}(a).
Then, along paths $b_1$ and $b_2$, 
the order parameter receives the transformation by 
the following group elements:
\begin{eqnarray}
b_1&:&  
\exp \left[{2i \over N} \left(\theta + {\pi \over 2}\right) \diag (1,1,\cdots,1)\right] 
\in U(1)_{\rm A} ,\quad \quad
-{\pi \over 2} \leq \theta \leq {\pi \over 2}, \non
b_2&:&  \omega_N \exp \left[{ 2i \over N} \left(\theta-{\pi\over 2}\right) \diag
       (N-1,-1,\cdots,-1)\right]  \in SU(N)_{\rm L-R}, \quad 
{\pi \over 2} \leq \theta \leq {3 \over 2}\pi .
\end{eqnarray}
Only the $U(1)_{\rm A}$ phase is rotated along path $b_1$,  
while only the $SU(N)_{\rm L-R}$ transformation 
is performed along path $b_2$. 
This configuration was discussed in Ref.~\cite{Balachandran:2002je}.
A numerical solution of the non-Abelian axial vortex with a fractional domain wall is shown in Fig.~\ref{fig:na}. 
However, note that
the vortex is pulled by the tension of the domain wall 
and consequently this configuration is not static~\footnote{
In general, it is not easy to obtain numerical solutions for non-static configurations. 
Here, we utilize a dynamical solution given 
in Fig.~\ref{fig:three_dw_junction2}, 
which we will show as an example of disintegration of an Abelian axial vortex in Sec.\ref{sec:instability}.
Let us look at the solution
at $t=20$ which is the panel surrounded by red line in Fig.~\ref{fig:three_dw_junction2}.
We trim Fig.~\ref{fig:three_dw_junction2} and pick up a local region $x\in(-25,-5)$, $y\in(-10,10)$, which
results in Fig.~\ref{fig:na}. 
Fig.~\ref{fig:na} is a snap shot at an instant and the non-Abelian axial vortex is continuously pulled to the left.}.
From Fig.~\ref{fig:na}, one can see that the domain wall interpolates $\varphi_{\rm A} = 2\pi/3$ and $\varphi_{\rm A} = 4\pi/3$ which ends on the non-Abelian axial vortex of $\phi_3$.

For $N=3$, there is another kind of junctions, 
called an M$_2$ 
non-Abelian vortex \cite{Balachandran:2005ev,Eto:2013hoa}. It takes the form 
\begin{eqnarray} 
\Sigma 
\xrightarrow[]{r\to\infty} 
v\,{\rm diag}\left(1,e^{i\theta},e^{i\theta}\right)
=v\,e^{i\frac{2\theta}{3}} {\rm
diag}\left(e^{-i\frac{2\theta}{3}},e^{i\frac{\theta}{3}},e^{i\frac{\theta}{3}}\right), 
\end{eqnarray}
in the absence of the instanton-induced potential. 
In the presence of the instanton-induced potential, 
the $U(1)_{\rm A}$ phase rotates by $-2\pi/3$. 
Therefore, two axial domain walls are attached 
as illustrated in Fig.~\ref{fig:NAvortex-wall}(b) \cite{Balachandran:2002je}.

\section{Instability of domain wall junctions}\label{sec:instability}
\begin{figure}[ht]
\begin{center}
\includegraphics[width=16cm]{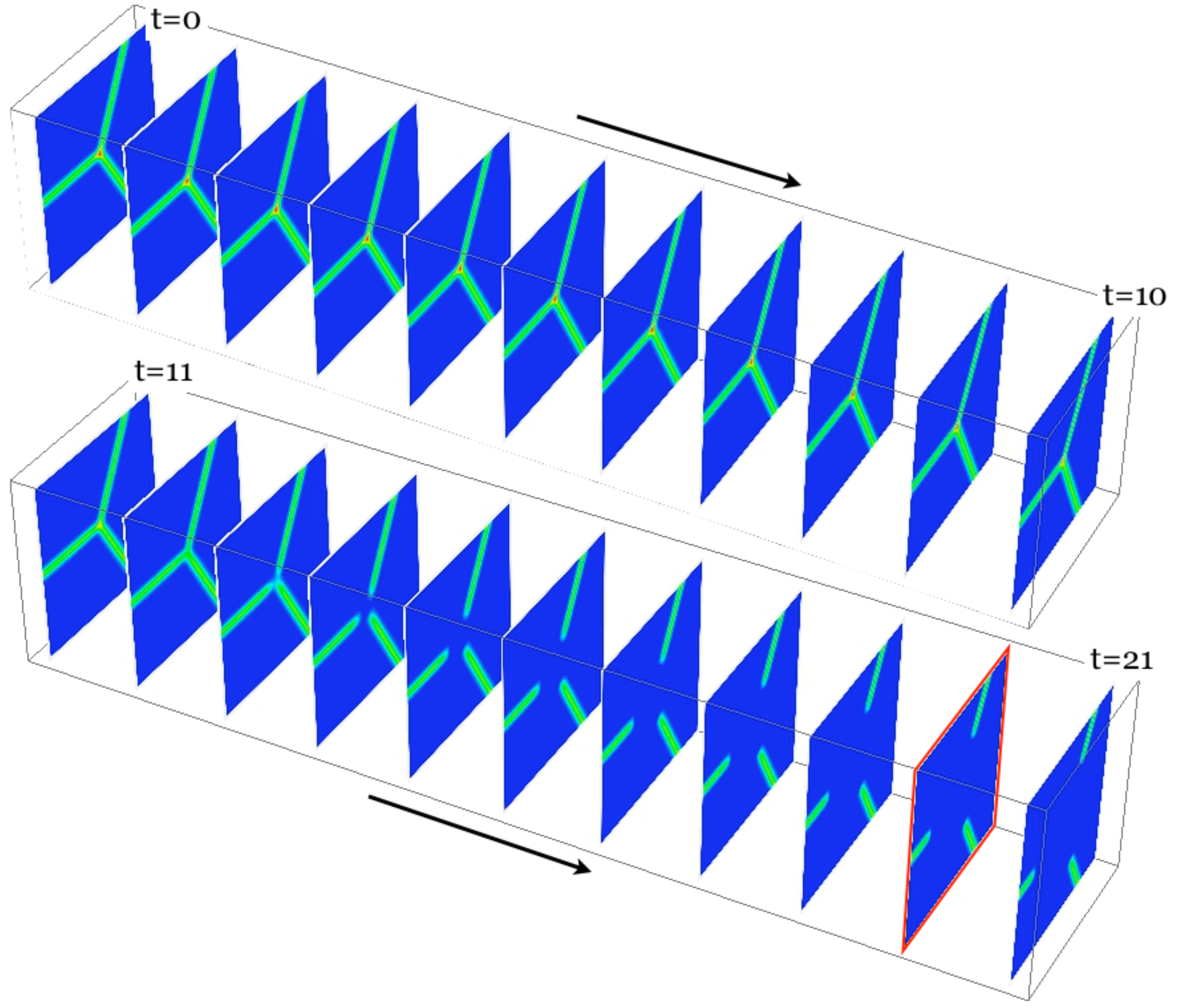}
\caption{
A decay sequence of a three-pronged fractional axial domain wall junction for $N=3$ case. 
The potential energy densities are plotted.
We put tiny random noise around a junction point at $t=0$, which disunites 
the Abelian axial vortex. 
The snapshots are taken from $t=0$ to $t=21$ with $\Delta t=1$. 
The parameter is $\tau \simeq 0.5$ ($\lambda_1=\lambda_2=1$ and $C/\mu = 1$).
The plotted region is $x\in(-20,20)$, $y\in(-20,20)$
in units of $m_1^{-1}$.
We have taken the configuration in Fig.~\ref{fig:na} from 
the panel outlined with a red square at $t=20$.
The colors correspond to the height of the potential energy density as
blue (low energy) $\to$ green $\to$ yellow $\to$ red (high energy).
}
\label{fig:three_dw_junction2}
\end{center}
\end{figure}

The domain wall junctions shown in Fig.~\ref{fig:junction}
or in Fig.~\ref{fig:NAvortex-wall}(b) 
were considered to be stable \cite{Balachandran:2002je}.
However, from now on we will show that 
they are in fact unstable. 
Here we will study the $N=3$ model for simplicity, but it is straightforward to extend it to generic $N$.
A junction of three fractional axial domain walls 
decays into a set of three fractional axial domain walls, each of which
is edged by non-Abelian axial vortices.
An Abelian axial vortex is attached 
by no domain walls
in the absence of the instanton-induced potential, 
as discussed in Sect.~\ref{eq:axial_vortex}. 
Nevertheless,  
it can be separated into three non-Abelian axial vortices as in Eq.~(\ref{eq:U(1)A-decay})
without binding force at the leading order \cite{Nakano:2007dq}. 
Since the axial anomaly is always present in reality, 
the configuration of 
a single Abelian axial vortex 
attached by three domain walls is unstable and it decays
as shown in Fig.~\ref{fig:three_dw_junction2},
since each non-Abelian axial vortex is pulled by the tension of 
a fractional axial domain wall.  
The $\U(1)_{\rm A}$ phase changes by $2\pi/3$ 
around each non-Abelian axial vortices
attached by fractional axial domain walls. 

\begin{figure}[t]
\begin{center}
\includegraphics[height=6cm]{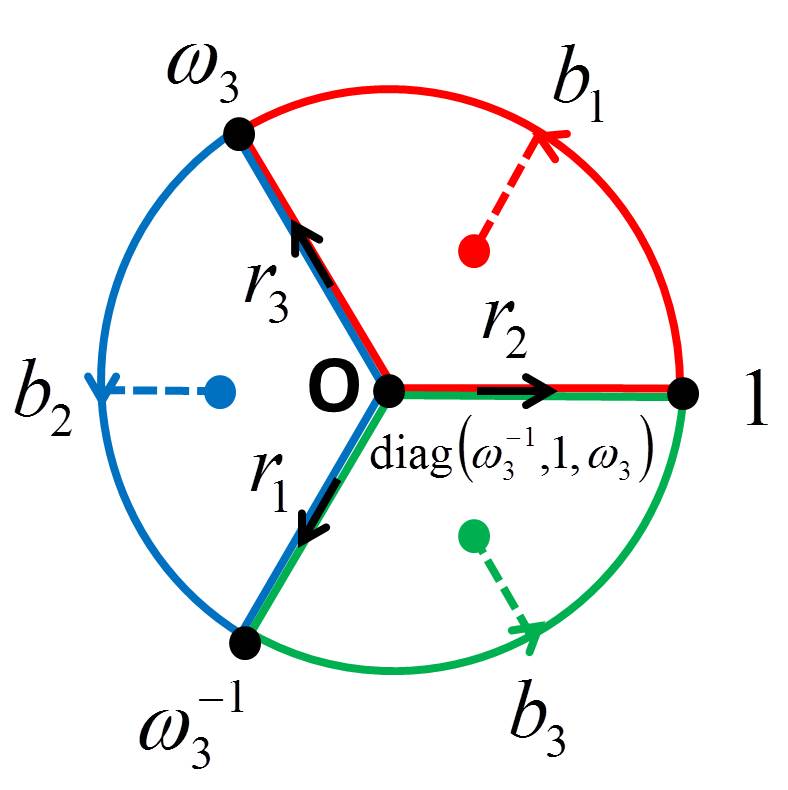}
\caption{
Classical decay of an axial domain wall junction.
See text for explanation. 
}
\label{fig:wall-junction-decay}
\end{center}
\end{figure}

We show the detailed configuration of a decaying junction in
Fig.~\ref{fig:wall-junction-decay}.
The Abelian axial vortex initially located at the origin O decays 
into three non-Abelian axial vortices, denoted by 
the red, green, and blue dots. 
The three fractional axial domain walls 
 denoted by the red, blue, and green dotted lines 
initially separate 
$\Sig \sim {\bf 1}_3$ and $\omega_3 {\bf 1}_3$, $\omega_3 {\bf 1}_3$ and 
$\omega_3^{-1} {\bf 1}_3$, and $\omega_3^{-1} {\bf 1}_3$ and 
${\bf 1}_3$,  respectively. 
The red, blue, and green non-Abelian axial vortices 
are encircled by the paths
$b_1 - r_3 + r_2, 
 b_2 - r_1 + r_3,
 b_3 - r_2 + r_1$
respectively.
At the boundary of the spatial infinity, 
the $U(1)_{\rm A}$ phase is rotated by 
$\exp [i \theta \diag (1,1,1)]$ with the angle $\theta$ 
of the polar coordinates from the origin O.  
Therefore,  
the $U(1)_{\rm A}$ phase is rotated 
by $2\pi/3$ along each of the paths 
$b_1$, $b_2$ and $b_3$. 
Let us suppose that the three paths 
enclose the three configurations in Eq.~(\ref{eq:ansz_ngv}),
respectively. Then, we find that 
the transformations $g(r) \in SU(3)_{\rm L-R}$ occur 
along the paths $r_1$, $r_2$ and $r_3$ as 
\begin{eqnarray}
r_1: && g(r) 
= \exp [i u(r) \diag(0,-1,1)]
=\bigg\{\begin{array}{c}
  \diag (1,1,1), \quad r=0 \cr 
  \diag (1,\omega_3^{-1},\omega_3) , \quad r=\infty
\end{array}, \non
r_2: && g(r) 
= \exp [i u(r) \diag(1,0,-1)]
=\bigg\{\begin{array}{c}
  \diag (1,1,1), \quad r=0 \cr 
  \diag (\omega_3,1,\omega_3^{-1}) , \quad r=\infty
\end{array}, \label{eq:decay-path2}\\
r_3: && 
g(r) = \exp [i u(r) \diag(-1,1,0)]
=\bigg\{\begin{array}{c}
  \diag (1,1,1), \quad r=0 \cr 
  \diag (\omega_3^{-1},\omega_3,1) , \quad r=\infty
\end{array}, \nonumber
\end{eqnarray}
respectively, 
where  $u(r)$ is a monotonically increasing function  
with the boundary conditions 
$u(r=0)=0$ and $u(r=\infty)=2\pi/3$. 
We find that the origin O is consistently given by 
$\Sigma = v \diag (\omega_3^{-1},1,\omega_3)$.
From a symmetry, permutations of each component 
are equally possible. 
An M$_2$ non-Abelian axial vortex in 
Fig.~\ref{fig:NAvortex-wall}(b)  
also decays into two non-Abelian axial vortices for the same reason.  

The configurations studied here are topologically the same 
\cite{Eto:2013hoa}
with  a $U(1)_{\rm B}$ superfluid vortex 
broken into a set of three semi-superfluid non-Abelian vortices 
in dense QCD \cite{Balachandran:2005ev,Eto:2009kg,Eto:2009bh}.

Note that there is a sharp contrast to the axion strings. 
Though an axion string in the $N=3$ axion model 
also gets attached by three domain walls, 
the domain walls cannot tear off the axion string into three fractional strings \cite{Vilenkin:1994}.

Before closing this section, let us make a comment on the effects by quark masses.
The quark masses can be taken into account 
 in the effective Lagrangian (\ref{eq:lag_lsm}),  
as an additional term $
{\rm Tr}\left[M(\Sigma + \Sigma^\dagger)\right]
$ 
with $M \propto {\rm diag}(m_u,m_d,m_s)$.
In order to see the deformation of the potential, it is useful to use the restricted field given in Eq.~(\ref{eq:restrict}) again,
and one finds that the axial phase receives an additional potential $\sim v (m_u+m_d+m_s) \cos \varphi_{\rm A}$.
So the potential has two terms $\cos 3\varphi_{\rm A}$ and $\cos \varphi_{\rm A}$ in competition with each other.
When the quark masses are small enough to be neglected, the Abelian axial vortex is torn off by three 
domain walls. On the other hand, when the quark masses are large enough compared to the instanton-induced potential,
there is only one true ground state, so that the Abelian axial vortex cannot be separated into three non-Abelian axial
strings. The three domain walls are glued into one {\it fat} domain wall and it will attach to an Abelian axial vortex.
A detailed analysis, including numerical solutions, is given elsewhere \cite{Eto:2013hoa}.

\section{Quantum decay of axial domain walls}
We here discuss the quantum decay of fractional 
axial domain walls. 
Although this domain wall is classically stable, 
it turns out to be metastable if one takes into account
the quantum tunneling effect. 
Inside a fractional axial domain wall, quantum (or thermal) fluctuations
make holes,
which are edged with non-Abelian vortices.
If a hole exceeds the critical size, it expands, just as a leaf is eaten
by caterpillars, 
because of the tension of the domain wall.
Eventually, the domain wall disappears \cite{Bachas:1994ds}.
The energy of the domain walls mainly turns into to the radiated $\eta'$ mesons and pions. 

\begin{figure}[ht]
\begin{center}
\includegraphics[width=12cm]{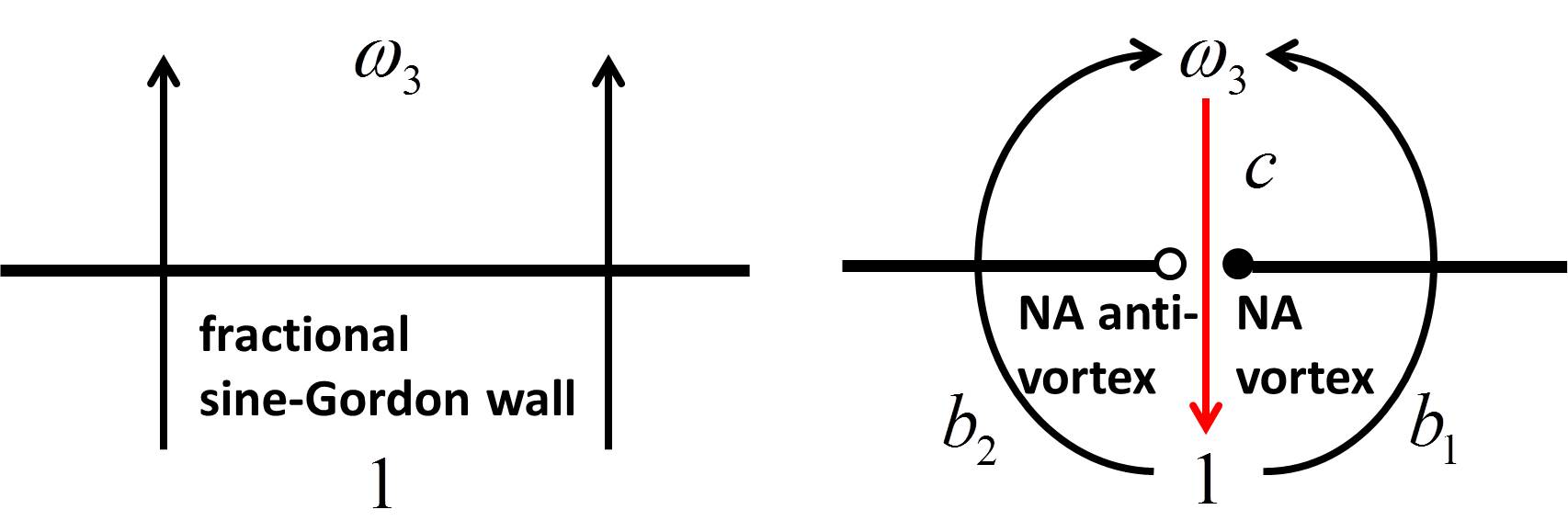}
\caption{Quantum decay of a fractional axial domain wall. 
A pair of a non-Abelian axial vortex and a non-Abelian 
axial anti-vortex is created. 
}
\label{fig:wall-decay}
\end{center}
\end{figure}

This should be contrasted with the $N>1$ axion model, where the
potential has the same periodicity $\varphi_{\rm A} \sim \varphi_{\rm A} +
2\pi/N $ and domain walls are stable. 
The difference comes from the fact that degenerate ground states 
in the case of chiral phase transition can be connected 
by a path in the $SU(N)_{\rm L-R}$ group 
without a potential, as explained above.
Let us first consider $d=2+1$ dimensions for simplicity. 
Suppose we have an axial domain wall interpolating 
between $\Sigma \sim {\bf 1}_N$ and 
$\Sigma \sim \omega_N {\bf 1}_N$ 
as in the left panel of Fig.~\ref{fig:wall-decay}. 
This wall can decay by creating 
path $c$ in the right panel of Fig.~\ref{fig:wall-decay}, 
along which the two ground states $1$ and $\omega_N$ 
are connected by 
\beq 
\omega_N\, \exp \left[ {i\over N} \left(\theta-{\pi\over 2}\right) \diag (N-1,-1,\cdots,-1)\right]  
 =
\bigg\{\begin{array}{c}
  \omega_N, \quad \theta = {\pi \over 2}\cr 
  1         , \quad  \theta ={3\over 2} \pi
\end{array}
\eeq
in the $SU(N)_{\rm L-R}$ group ($\pi/2 \le \theta \le 3\pi/2$).  
Here $\theta$ represents the angle from the black point. 
Then, one finds that the counterclockwise loop $b_1+c$ 
encloses a non-Abelian axial vortex 
of the type diag$(e^{i\theta},1,\cdots,1)$
(represented by the black point).
This is nothing but the configuration in Fig.~\ref{fig:NAvortex-wall}.  
The clockwise closed loop $-b_2+c$ also encloses a non-Abelian axial
vortex (denoted by a white point), 
which implies that it is an non-Abelian axial anti-vortex. 
Therefore, a hole bounded by a pair of  
a non-Abelian axial vortex and a non-Abelian axial anti-vortex
is created.  
When one deforms the path $b_1$ to $-c$ in Fig.~\ref{fig:wall-decay}, 
one must create a non-Abelian vortex, 
implying an energy barrier between these two paths.
Therefore, the domain wall is metastable. 

In $d=3+1$ dimensions, a 2D hole bounded by 
a closed non-Abelian axial vortex loop is created. 
Through this decay process, the domain wall energy turns into
radiation of the $U(N)_{\rm L-R+A}$ Nambu-Goldstone modes 
($\eta'$ mesons and pions).

The decay rate of axial domain walls can be calculated as follows \cite{Preskill:1992ck}.  
Once a hole is created on the integer axial wall, it will expand if the
size of this hole 
is larger than a critical value, and the axial domain wall decays. 
We calculate the quantum tunneling probability of this
process. 
Let $R$ be the initial radius of a hole created on the axial domain wall.
Then, the bounce action of this tunneling process is 
\begin{equation}
B = 4\pi R^2 T_{\rm v} - \frac{4}{3}\pi R^3 T_{\rm w},
\label{eq:decay-prob1}
\end{equation}
where $T_{U(N)_{\rm L-R+A}}$ and $T_{\rm w}$ are the tensions of the vortex and the
axial domain wall, given in Eqs.~(\ref{eq:tension-NAvor}) and (\ref{eq:tension-wall}), respectively.
The critical radius $R_{\rm c}$ is the one that minimizes this bounce
action, given by 
$
R_{\rm c} = 2T_{U(N)_{\rm L-R+A}}/T_{\rm w} . 
$ 
Thus, the decay rate is
\beq
P &\sim& e^{-B}\big|_{R=R_{\rm c}} = \exp\left(-
\frac{16\pi}{3}\frac{T_{U(N)_{\rm L-R+A}}^3}{T_{\rm
w}^2}\right)\\
&=& \exp\left(-\frac{N^2\pi^4v^2}{6 m_{\eta'}^2}\left(\log\frac{L}{\xi_{\rm na}}\right)^3\right).
\label{eq:decay-prob2}
\eeq

\section{Summary and Discussion} \label{sec:summary}
We have studied domain walls and vortices  in 
the broken phase of the chiral symmetry 
in QCD with $N$ flavors in the chiral limit.
In the absence of the axial anomaly, there exist stable 
Abelian axial vortices winding around the spontaneously broken 
$U(1)_{\rm A}$ symmetry
and non-Abelian axial vortices 
winding around both 
the $U(1)_{\rm A}$ 
and non-Abelian $SU(N)$ chiral symmetries. 
In the presence of the axial anomaly term, 
metastable domain walls are present and 
vortices cannot exist alone.  
Abelian axial vortices are attached by 
$N$ domain walls forming domain wall junctions, 
and a non-Abelian axial vortex is attached by 
a domain wall.  
We have argued 
that a domain wall junction can topologically decay 
 into $N$ non-Abelian vortices attached by 
domain walls  
implying its metastability, 
and simulated such a decay numerically. 
We have also shown that domain walls can decay 
quantum-mechanically by creating 
a hole bounded by a closed non-Abelian vortex. 

In order to study whether the 
domain wall problem exists, 
we have to estimate how many domain walls 
are created in the phase transition by the Kibble-Zurek mechanism \cite{Kibble:1976sj,Hindmarsh:1994re,Zurek:1985qw,Zurek:1996sj}.  
Since the chiral symmetry breaking is actually a crossover rather than a
phase transition, the estimation of the domain wall number density is not
straightforward. 
Then, the mechanism found in this paper would 
reduce the number of domain walls. 
Numerical simulation of the production and decay of 
domain walls remains as an important future problem. 
It would  also be interesting to study these processes in 
heavy-ion collisions. 

As described in the introduction, the same discussions 
in this paper hold for chiral symmetry breaking 
in high-density QCD \cite{Eto:2013hoa}.  
However, there is also a difference because of 
the color degrees of freedom in the symmetry breaking; 
In addition to the non-Abelian axial vortices discussed in this paper, 
there are also 
non-Abelian semi-superfluid vortices, 
which are color magnetic flux tubes 
\cite{Balachandran:2005ev,Nakano:2007dr,Nakano:2008dc,
Eto:2009kg,Eto:2009bh,Eto:2009tr,Hirono:2010gq}. 
The roles played by these flux tubes is an open question. 

\section*{Acknowledgements}

This work is supported in part by 
Grant-in-Aid for Scientific Research (Grants
No. 23740198 (M.E.),  No. 25400268 (M.N.)).
The work of 
Y.H. is partially supported by the Japan Society for the Promotion of
Science for Young Scientists and partially by the JSPS Strategic Young
Researcher Overseas Visits Program for Accelerating Brain Circulation (No.R2411). 
The work of M.N. is also supported in part by 
the ``Topological Quantum Phenomena'' 
Grant-in-Aid for Scientific Research 
on Innovative Areas (Grant No. 25103720)  
from the Ministry of Education, Culture, Sports, Science and Technology (MEXT) of Japan.


\newcommand{\J}[4]{{\sl #1} {\bf #2} (#3) #4}
\newcommand{\andJ}[3]{{\bf #1} (#2) #3}
\newcommand{\AP}{Ann.\ Phys.\ (N.Y.)}
\newcommand{\MPL}{Mod.\ Phys.\ Lett.}
\newcommand{\NP}{Nucl.\ Phys.}
\newcommand{\PL}{Phys.\ Lett.}
\newcommand{\PR}{ Phys.\ Rev.}
\newcommand{\PRL}{Phys.\ Rev.\ Lett.}
\newcommand{\PTP}{Prog.\ Theor.\ Phys.}
\newcommand{\hep}[1]{{\tt hep-th/{#1}}}
\bibliographystyle{ptephy}
\bibliography{dense-QCD}

\end{document}